\documentclass[twocolumn,letterpaper,superscriptaddress,showpacs,amsmath,amssymb,floatfix,amstex]{revtex4}
\usepackage{graphicx,bbm}

\usepackage{color}

\usepackage[slovene,english]{babel} 
\usepackage{amsthm}
\usepackage{dsfont}

\newcommand{\eref}[1]{(\ref{#1})}
\newcommand{\aref}[1]{Appendix \ref{#1}}
\newcommand{\fref}[1]{Figure \ref{#1}}

\newcommand{\mc}[1]{\mathcal{#1}}

\newcommand{\ul}[1]{\underline{#1}}

\newcommand{\trb}[1]{ {\rm tr}\left( #1\right) }
\newcommand{\tr}{ {\rm tr} }
\newcommand{\bra}[1]{ \left \langle #1\right | }
\newcommand{\bbra}[1]{ \left \langle #1\middle|\! \right | }

\newcommand{\ket}[1]{ \left | #1\right \rangle}
\newcommand{\kket}[1]{ \left |\! \middle| #1\right \rangle}
\newcommand{\braket}[2]{ \left \langle #1\middle| #2\right \rangle}
\newcommand{\brakket}[2]{ \left \langle #1\middle|\!\middle| #2\right \rangle}

\newcommand{\dd}{ {\rm d} }
\newcommand{\ddt}{\frac{\dd}{\dd t} }
\newcommand{\ii}{ {\rm i} }

\newcommand{\ee}{ {\rm e} }

\newcommand{\ave}[1]{{\langle #1\rangle}}

\def\tr{{\,{\rm tr}\,}}

\def\sy{\sigma^{\rm y}}
\def\sz{\sigma^{\rm z}}
\def\sx{\sigma^{\rm x}}

\begin{document}
\title{Continuous phase-space methods on discrete phase spaces}%{Discretisations of $SU(N)$ quasiprobability distributions}
 
\author{Bojan \v{Z}unkovi\v{c}}
\affiliation{Departmento la Fisico, Universidad de Chile, Santiago, Chile}
\affiliation{Scuola Internazionale Superiore di Studi Avanzati, Trieste, Italy}

%\address{Scuola Internazionale Superiore di Studi Avanzati, Trieste, Italy\\$^{2}$Departmento la Fisico, Universidad de Chile, Santiago, Chile}

\date{\today}

\begin{abstract}
We show that discrete quasiprobability distributions defined via the discrete Heisenberg-Weyl group can be obtained as discretizations of the continuous $SU(N)$ quasiprobability distributions. This is done by identifying the phase-point operators with the continuous quantisation kernels evaluated at special points of the phase space. As an application we discuss the positive-$P$ function and show that its discretization can be used to treat the problem of diverging trajectories. We study the dissipative long-range transverse-field Ising chain and show that the long-time dynamics of local observables is well described by a semiclassical approximation of the interactions.
\end{abstract}

\maketitle
\section{Introduction}Phase-space representations of quantum mechanics, usually called quasiprobability distributions,  provide a natural language for the quantum-classical correspondence of non-relativistic quantum mechanics. They were first developed in the context of harmonic oscillators \cite{HOS+84} and extensively applied in quantum optics \cite{GZ04, Lee95}. It has been realised that the quantum-classical correspondence is intimately related to  symmetry properties of the underlying classical phase-space. This led to an axiomatic approach to the phase-space formulation of quantum mechanics known as Stratonovich-Moyal-Weyl correspondence \cite{KC09, Ber75}. In order to apply the ideas of Stratonovich and Berezin to finite dimensional Hilbert spaces (e.g. spin systems) two main research directions are followed. First is to construct a continuous phase space with a natural continuous symmetry of the system (e.g. SU(2) for spin systems) \cite{BM99, VG89, Agr81, KG10, TN11}.  In  many cases this permits to rewrite the evolution equations in terms of stochastic differential equations \cite{ZC07, Pol03} or even ordinary differential equations \cite{ZC07, DP15} that can be efficiently simulated.  The second direction is to exploit the discrete nature of the system and to define phase-space distributions on the basis of the discrete Heisenberg-Weyl group \cite{Woo86, GP92, LP98, Woo04, Vou04, KM05, GHW05, RMG05, CCF+13}. This formulation enables a simpler representation of quantum states and has been applied in quantum information \cite{Gro06, Fer10} and quantum tomography \cite{TDB+10, MRG05}.

We show that the two approaches to formulate phase-space quasiprobability distributions on finite dimensional Hilbert spaces are equivalent, in the sense that discrete distributions can be obtained by evaluating the continuous ones at special points of the phase space. This identification provides a formal justification of Monte-Carlo methods on finite dimensional phase spaces proposed in \cite{SPR15,SPR15a}, their systematic expansion beyond the semiclassical (truncated Wigner) approximation \cite{Pol03, Pol10} and extension to open quantum systems. Hence, the revealed relation between continuous and discrete phase spaces should be relevant to further development of contemporary phase-space methods for simulation of long-range many-body quantum systems. A similar discretisation was obtained in case of the $SU(2)$ group \cite{SS00, AW00} by contracting a continuous SU(2) kernel to a discrete one. Here we take the opposite route and extend the discrete kernel to the continuous one for any dimension of the Hilbert space $N$.

In order to prove the main result of the paper we first review the basic properties of the discrete and $SU(N)$ quasiprobability distributions on $N$-dimensional Hilbert space.
\subsection{Discrete quasiprobability distributions}A one parameter family of discrete Weyl symbols for an operator $A$ acting on a $N$-dimensional Hilbert space is usually defined as 
\begin{align}
W_{A}^{(s)}(\alpha,\beta)=\tr{A\Delta_{\alpha,\beta}^{(s)}},
\label{eq:discrete Weyl}
\end{align}
with the parameter $s$ denoting the ordering of the distribution; $s=1,0,-1$ for normal, symmetric, and anti-normal ordering, respectively. The phase-point operators $\Delta_{\alpha,\beta}^{(s)}$, $\alpha,\beta=0,1,2,\ldots N-1$ satisfy the following properties \cite{LP98, RMG05, KM05}:
\begin{enumerate}
\item[1a)] Hermicity: $(\Delta_{\alpha,\beta}^{(s)})^\dag=\Delta_{\alpha,\beta}^{(s)}$,
\item[2a)] Normalization: $\tr{\Delta_{\alpha,\beta}^{(s)}}=1$,
\item[3a)] Covariance: $\Delta_{\alpha-\mu,\beta-\nu}^{(s)}=T_{\mu,\nu}\Delta_{\alpha,\beta}^{(s)}T^\dag_{\mu,\nu}$, where $T_{\mu,\nu}$ denotes a unitary irrep of the discrete Heisenberg-Weyl group of order $N$,
\item[4a)] Traciality: $\tr{\Delta_{\alpha,\beta}^{(s)}\Delta_{\alpha',\beta'}^{(-s)}}=N\delta_{\alpha,\alpha'}\delta_{\beta,\beta'}$.
\end{enumerate}
The quasiprobability distribution is  given by the Weyl symbol of the density matrix. For each $\Delta_{\alpha,\beta}$ satisfying the properties 1a--4a the Weyl symbol \eref{eq:discrete Weyl} defines a one-to-one mapping between operators on a finite dimensional Hilbert space and functions on a discrete phase space. 
%The kernel for an $N$-dimensional Hilbert space can be explicitly written as 
%\begin{align}
%\Delta_{\alpha,\beta}=\frac{1}{N}\sum_{m,n=-\lfloor N/2\rfloor}^{\lfloor N/2\rfloor}\omega(\alpha m-\beta n)T_{m,n},
%\end{align}
%with $\omega(m)=\exp(2\pi\ii/N)$. 
The explicit form of the phase-point operators is not relevant for our purpose and will be omitted; we refer the interested reader to references \cite{Woo86, GP92, KM05} (for a particular example see Section \ref{sec:geom}). The kernel $\Delta_{\alpha,\beta}^{(s)}$ can be rewritten in the following useful form \cite{KM05}
\begin{align}
\Delta_{\alpha,\beta}^{(s)}=T_{\alpha,\beta}\Delta_{0,0}^{(s)}T^\dag_{\alpha,\beta}.
\label{eq:decomp1}
\end{align}
The decomposition \eref{eq:decomp1} is valid for any unitary realisation $T_{\alpha,\beta}$ of the discrete Heisenberg-Weyl group of order $N$. The constant matrix $\Delta_{0,0}^{(s)}$ is determined by demanding that for $s=-1$ the quasiprobability distribution represents the discrete Q function. This gives the boundary condition
\begin{align}
\Delta_{\alpha,\beta}^{(-1)}=T_{\alpha,\beta}\ket{0}\bra{0}T_{\alpha,\beta}^\dag,
\end{align}
where $\ket{0}$ denotes a chosen vacuum state. Other $\Delta^{(s)}_{0,0}$ can be determined by using the traciality condition \cite{RMG05}. 
%The identity \eref{eq:decomp1} shall be extensively used in the proof of the main result in the next section. 
Phase-point operators satisfying the properties 1a-4a can be defined and calculated for any Hilbert space dimension $N$. If in addition $N$ is a prime number or a power of a prime number the quasiprobability distribution is defined on a discrete phase space with a well defined geometry \cite{Woo86, Vou04, Hir98}. For such systems (sometimes called Galois quantum systems) phase-space methods based on discrete symplectic transformations were developed that are similar to the ones used for harmonic oscillator systems \cite{Vou04}. Phase spaces with non-prime number dimensions $N$ do not have a well defined geometry. As a consequence the standard construction of the mutually unbiased basis does not work, leaving this fundamental and interesting problem unsolved in the Hilbert spaces of non-prime number dimension. In addition, on even dimensional phase spaces the discrete kernels are not uniquely determined \cite{LP98, Fer10}. This, however, does not affect our results.
\subsection{$SU(N)$ quasiprobability distributions}The $SU(N)$ quasiprobability distributions on a manifold $\mathcal{M}$ are constructed in a similar manner as the discrete quasiprobability distributions through the quantisation kernel $\Delta^{(s)}(\Omega)$. The Weyl symbol is given by 
\begin{align}
W^{(s)}_A(\Omega)=\tr{A\Delta^{(s)}(\Omega)},
\label{eq:cont Weyl}
\end{align}
with the parameter $s=-1,0,1$ denoting the normal, symmetric, and anti-normal ordering. The quantisation kernel has to satisfy similar properties as in the discrete case, namely \cite{KC09}
\begin{enumerate}
\item[1b)] Hermicity: $(\Delta^{(s)}(\Omega))^\dag=\Delta^{(s)}(\Omega)$,
\item[2b)] Normalization: $\tr{\Delta^{(s)}(\Omega)}=1$,
\item[3b)] Covariance: $\Delta^{(s)}_{g^{-1}\circ\Omega}=\Lambda(g) \Delta^{(s)}(\Omega)\Lambda^\dag(g)$, where $\Lambda(g)$ denotes a unitary irrep of a coset element $g\in\mathcal{G}=SU(N)/U(N-1$),
\item[4b)] Traciality: $\int_{\mathcal{M}} \dd \mu(\Omega)\trb{\Delta^{(s)}(\Omega)\Delta^{(-s)}(\Omega')}f^{(s)}(\Omega)=f^{(s)}(\Omega')$, where $\dd \mu(\Omega)$ is the invariant measure, $f^{(s)}(\Omega)$ denotes a differentiable $s$-ordered function on the manifold $\mathcal{M}$ being isomorphic to $\mathcal{G}$.
\end{enumerate}
For each quantisation kernel $\Delta^{(s)}(\Omega)$ satisfying the properties 1b-4b the Weyl symbol \eref{eq:cont Weyl} represents a one-to-one mapping between operators on the $N$-dimensional Hilbert space and smooth functions on the classical manifold $\mathcal{M}$. There are several constructions of the quantisation kernel for the $SU(N)$ group \cite{BM99, KG10, TN11}. Here we focus on the fundamental representation for which the most explicit expression for the quantisation kernel was given in \cite{TN11}. The explicit form of the $SU(N)$ kernel shall be omitted; interested reader is referred to \cite{BM99, KG10, TN11} (for a particular example see Section \ref{sec:geom}). As in the discrete case all realisations of the $SU(N)$ group admit the following decomposition of the kernel \cite{KG10}
\begin{align}
\Delta^{(s)}(\Omega)=\Lambda(\Omega)D^{(s)}\Lambda(\Omega)^\dag,
\end{align}
where $D^{(s)}$ is a constant diagonal matrix containing the essential information about the quasiprobability distribution. It can be determined by demanding that for $s=-1$ the usual Q function should be recovered, namely
\begin{align}
\Delta^{(-1)}(\Omega)=\Lambda(\Omega) \ket{0}\bra{0}\Lambda(\Omega)^\dag.
\end{align}
All other $D^{(s)}$ are determined by the traciality condition \cite{TN11, KG10}.

\section{Construction of the $SU(N)$ kernel from the phase-point operators}
\label{sec:geom}
In the following we shall prove that continuous $SU(N)$ quasiprobability distributions for the fundamental representation of the $SU(N)$ group can be constructed from the discrete phase-point operators. We start by realising that the discrete Heisenberg-Weyl group of order $N$ is a subgroup of the coset $\mathcal{G}$. Hence, the realisations $T_{\alpha,\beta}$ associated to the $N$-dimensional Hilbert space can be seen as fundamental realisations of particular group elements $\Omega_{\alpha,\beta}\in\mathcal{G}$, namely $T_{\alpha,\beta}=\Lambda(\Omega_{\alpha,\beta})$, suggesting that the phase-point operators can be regarded as the $SU(N)$ kernels evaluated at phase-space points $\Omega_{\alpha,\beta}$.  This is immediately clear for the Q function, however, it can be shown for any $s$ without knowing the precise form of the kernels $\Delta^{(s)}_{\alpha,\beta}$ and $\Delta^{(s)}(\Omega)$.

The discrete phase-point operators can be extended to the $SU(N)$ group by defining
\begin{align}
\Delta^{(s)}_{\alpha,\beta}(\Omega)=\Lambda(\Omega)\Delta^{(s)}_{\alpha,\beta}\Lambda^\dag(\Omega).
\label{eq:gkernel}
\end{align}
The extended Weyl symbol for an operator $A$ is then defined as 
\begin{align}
W^{(s)}_{A}(\alpha,\beta,\Omega)=\trb{A\Delta^{(s)}_{\alpha,\beta}(\Omega)}.
\label{eq:general symbol}
\end{align}
In the following we shall show that for any fixed $\Omega$ the matrix $W_{A}^{(s)}(\alpha,\beta,\Omega)$ represents a discrete Weyl symbol and that for any fixed pair $\alpha,\beta$ the function $W_{A}^{(s)}(\alpha,\beta,\Omega)$ represents a continuous Weyl symbol. The properties 1a, 2a, and 4a of the phase-point operators are trivially satisfied. The covariance property 3a follows by using a rotated realisation of the discrete Heisenberg-Weyl group, namely $T_{\alpha,\beta}(\Omega)=\Lambda(\Omega)T_{\alpha,\beta}\Lambda^\dag(\Omega)$. Further, the properties 1b-3b of the continuous quantisation kernel are evident. Finally, the continuous traciality property 4b follows from the invariance of the Haar measure and from the fact that $\Delta_{\alpha,\beta}(\Omega)$ are also discrete phase-point operators for any but fixed $\Omega\in\mathcal{G}$. By using these two properties any operator $A$ acting on the Hilbert space can be decomposed as
\begin{align}
\label{eq:trac1}
A&=\sum_{\alpha,\beta}\tr(\Delta_{\alpha,\beta}^{(-s)}(\Omega) A)\Delta^{(s)}_{\alpha,\beta}(\Omega)\\ \nonumber
&=\frac{1}{\Omega_N}\int_{\mathcal{M}} \dd\mu(\Omega) \sum_{\alpha,\beta}\tr(\Delta_{\alpha,\beta}^{(-s)}(\Omega) A)\Delta_{\alpha,\beta}^{(s)}(\Omega)\\ \nonumber
&=\frac{N^2}{\Omega_N}\int_{\mathcal{M}} \dd\mu(\Omega) \tr(\Delta_{\alpha,\beta}^{(-s)}(\Omega) A)\Delta_{\alpha,\beta}^{(s)}(\Omega),
\end{align}
where $\Omega_N$ denotes the invariant volume of the phase space $\mathcal{M}$.
From above equality \eref{eq:trac1} the continuous traciality condition follows
\begin{align}
f^{(s)}_{\alpha,\beta}(\Omega)&=\tr(F\Delta^{(-s)}_{\alpha,\beta}(\Omega))\\ \nonumber
&=\frac{N^2}{\Omega_N}\int_{\mathcal{M}}\dd \mu(\Omega) f^{(s)}_{\alpha,\beta}(\Omega')\trb{\Delta^{(s)}_{\alpha,\beta}(\Omega') \Delta^{(-s)}_{\alpha,\beta}(\Omega)}.
\end{align}
Hence, the kernel $\Delta_{\alpha,\beta}(\Omega)$ can be used to generate a discrete or a continuous phase space. 

\begin{figure}[h!!]
\vskip 0.5truecm
\includegraphics[width=0.25\textwidth]{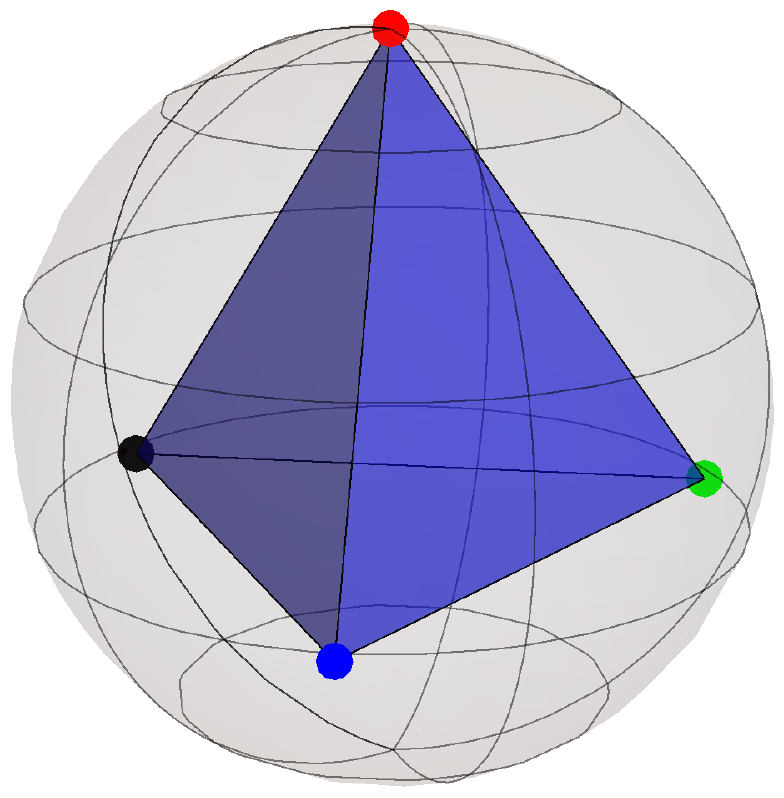}
\caption{Points in the continuous $SU(2)$ phase space corresponding to a particular choice of phase-point operators \eref{eqa:woot}. The red, green, blue and black points correspond to $z_{0,0},~z_{0,1},~z_{1,0}$ and $z_{1,1}$, respectively. Changing $\varphi$ rotates the discrete phase-space points around the $z$ (vertical) axis.}
\label{fig2}
\end{figure}

As an example we consider a spin 1/2 case, where the phase space is a sphere, which can be useful to gain some intuition on the mapping between the discrete to the continuous phase space.  We start by choosing the following phase-point operators
\begin{widetext}
\begin{align}
\label{eqa:woot}
A_{0,0}=\left(
\begin{array}{cc}
 \frac{1}{2} \left(1+\sqrt{3}\right) & 0 \\
 0 & \frac{1}{2} \left(1-\sqrt{3}\right) \\
\end{array}
\right),\quad A_{0,1}=\left(
\begin{array}{cc}
 \frac{1}{6} \left(3-\sqrt{3}\right) & \sqrt{\frac{2}{3}} \ee^{\ii\varphi}\\
 \sqrt{\frac{2}{3}} \ee^{-\ii\varphi}& \frac{1}{6} \left(3+\sqrt{3}\right) \\
\end{array}
\right),\\ \nonumber
A_{1,0}=\left(
\begin{array}{cc}
 \frac{1}{6} \left(3-\sqrt{3}\right) & (-1)^{2/3} \sqrt{\frac{2}{3}} \ee^{\ii\varphi}\\
 -\sqrt[3]{-1} \sqrt{\frac{2}{3}} \ee^{-\ii\varphi}& \frac{1}{6} \left(3+\sqrt{3}\right) \\
\end{array}
\right),\quad
A_{1,1}=\left(
\begin{array}{cc}
 \frac{1}{6} \left(3-\sqrt{3}\right) & -\sqrt[3]{-1} \sqrt{\frac{2}{3}} \ee^{\ii\varphi}\\
 (-1)^{2/3} \sqrt{\frac{2}{3}} \ee^{-\ii\varphi}& \frac{1}{6} \left(3+\sqrt{3}\right) \\
\end{array}
\right).
\end{align}
\end{widetext}
Using the $SU(2)$ kernel for the Wigner function 
\begin{align}
\label{eqa:ker}
\Delta^{(0)}(z)=\left(
\begin{array}{cc}
 \frac{\sqrt{3} \left(1-\left| z \right| ^2\right)}{2 \left(\left| z \right|
   ^2+1\right)}+\frac{1}{2} & \frac{\sqrt{3} z }{\left| z \right| ^2+1} \\
 \frac{\sqrt{3} \bar{z }}{\left| z \right| ^2+1} & \frac{\sqrt{3} \left(\left|
   z \right| ^2-1\right)}{2 \left(\left| z \right| ^2+1\right)}+\frac{1}{2} \\
\end{array}
\right)
\end{align}
 one can easily check that the phase-point operators \eref{eqa:woot} correspond to the continuous kernel \eref{eqa:ker} evaluated at special points of the phase space, namely $A_{i,j}=\Delta^{(0)}(z_{i,j})$ with $z_{0,0}=0$, $z_{0,1}=\sqrt{2}\exp\ii\varphi$, $z_{1,0}=(-1)^{2/3}\sqrt{2}\exp\ii\varphi$ and $z_{1,1}=-(-1)^{2/3}\sqrt{2}\exp\ii\varphi$. Applying the stereographic projection $z=\tan(\theta/2)\exp(\ii \phi)$ we find that the points corresponding to the  phase-point operators form vertices of a regular tetrahedron embedded in the continuous $SU(2)$ phase space (see \fref{fig2}). In \aref{App3} we use a slightly more general discrete phase space points, namely we also allow the rotation around $y$ axis of the phase space.

We have established a connection between discrete quasiprobability distributions and $SU(N)$ quasiprobability distributions on finite-dimensional Hilbert spaces. This identification enables to combine sampling of the initial state in the discrete phase-space and continuous simulation in the $SU(N)$ phase space \cite{SPR15}. Further, continuous phase-space methods enable a systematic expansion in the quantum noise parameter \cite{Pol03, ZC07} and extension to the dissipative models \cite{ZC07}. However, in order to use the phase-space methods one needs to find a differential representation for the operators on the local Hilbert space. This is in general a difficult task, which was accomplished in for any $SU(N)$ group only in the $s=1$ (P representation)  case \cite{BD08}, and for any $s$ only in the two dimensional case \cite{ZC07} (i.e. for the SU(2) group). Once the equations of motion in the phase space are obtained the main obstacles for their efficient simulation are divergence of stochastic trajectories and negativity of the diffusion kernel. The later can be avoided by using the  positive-$P$ representation (positive-$P$ function), whereas the former can be to some extend circumvented by exploiting a particular gauge freedom \cite{DD02}.

\section{Discrete positive-P distribution, diverging trajectories and noise reduction} In the rest of the paper we discuss the applicability of the positive-$P$ function on discrete phase spaces and propose a discrete phase-space method to treat the problem of diverging trajectories. The normalised positive-$P$ function is usually defined through the off-diagonal coherent state expansion of the density matrix \cite{BD08}
\begin{align}
\rho=\int_{\mathcal{M}}\dd \mu(\Omega)\dd \mu(\Omega')P^+(\Omega,\Omega')\frac{\ket{\Omega}\bra{\Omega'}}{\braket{\Omega'}{\Omega}}.
%\rho=\int\dd \mu(\Omega)\dd \mu(\Omega')P^+(\Omega,\Omega')\kket{\Omega}\bbra{\Omega'},
\label{eq:P+}
\end{align}
where $\ket{\Omega}:=\Lambda(\Omega)\ket{0}$ denotes a normalized coherent state. We can write above equation \eref{eq:P+} in terms of the quantisation kernel \eref{eq:gkernel} as 
\begin{align}
%\rho=\int\dd \mu(\Omega)\dd \mu(\Omega')\sum_{\alpha,\beta}P^{(s)}_{\alpha,\beta}(\Omega,\Omega')\Delta^{(-s)}_{\alpha,\beta}(\Omega,\Omega'),
\rho=\int_{\mathcal{M}}\dd \mu(\Omega)\dd \mu(\Omega')P^{(s)}(\Omega,\Omega')\Delta^{(-s)}(\Omega,\Omega'),
\label{eq:+D}
\end{align}
with the kernel
\begin{align}
\Delta^{(s)}(\Omega,\Omega')=\frac{\Lambda(\Omega)\Delta^{(s)}\Lambda^\dag(\Omega')}{\trb{\Lambda(\Omega)\Delta^{(s)}\Lambda^\dag(\Omega')}}.
\label{eq:+kernel}
\end{align}
By choosing $s=1$  in equation \eref{eq:+D} we obtain the density matrix expansion given in \eref{eq:P+} \footnote{In the cases $s=0,-1$ the kernel \eref{eq:+kernel} seizes to have appropriate analyticity properties with respect to the phase-space variables, hence, the positive-P distributions can not be generalised to $s=0,~-1$ cases in a straightforward way.}. The positive-P function $P^{(+1)}(\Omega,\Omega')$ always exists and can be expressed in terms of the discrete phase-space values as
\begin{align}
\label{eq:gen dis prob}
P^{(+1)}(\Omega,\Omega')&=P^{(+1)}(\Omega)\delta(\Omega_{\alpha,\beta}^{-1}\Omega)\delta(\Omega'),
\end{align}
where $P^{(+1)}(\Omega_{\alpha,\beta})=\trb{\rho\Delta_{\alpha,\beta}^{(+1)}}$ and $\delta(\Omega)$ is the delta function for the measure $\mu(\Omega)$. Above equation \eref{eq:gen dis prob} means that the discrete P distribution can be interpreted as a continuous positive-P distribution. Hence, the discrete-P representation can be used to sample the initial condition in simulations of the positive P function. 
Importance of the initial condition sampling and a comparison of continuous and discrete distributions was discussed in \cite{SPR15}. However, the derivation in \cite{SPR15} is valid only for Wigner functions of spin 1/2 systems and is based on a product-probability assumption, which results in a truncated Wigner-type approximation not permitting a systematic expansion or application to open quantum systems. The approach presented here is valid for any $N$, any s-ordered quasiprobability distribution and the generalised positive-P function, enables a systematic expansion in the noise terms, and is applicable to open quantum systems. 

Using the phase-space correspondence one can express the action of the elements of the $su(N)$ algebra on the density matrix by using only first derivatives with respect to the phase-space variables \cite{BD08}. Hence, any evolution equation for the density matrix which contains at most linear elements in the $su(N)$ generators can be expressed as an ordinary differential equation on the extended phase space. On the other hand, any evolution equation which contains at most quadratic elements in the $su(N)$ generators can be expressed as a stochastic differential equation on the extended phase space. Stochastic terms arise from the interaction between different sites of a many-body system or dissipation and make the simulation of evolution equations inefficient for longer times. One procedure to reduce the noise is to add a gauge or to experiment with different decompositions of the diffusion matrix \cite{DD02}. Another possibility is to enlarge the local Hilbert space with dimension $N$ by including $k$ nearest neighbours. Inside the block the evolution in the $SU(N^k)$ generalised phase space is described by ordinary differential equations, whereas the interaction between the blocks is still treated stochastically. The decomposition into larger blocks is still exact but reduces the number of stochastic terms at the expense of enlarging the local phase space \footnote{A similar method was proposed in \cite{DP15}, where the interaction between the blocks was treated in the first order approximation (TWA). In our case the evolution equation is still exact and the interaction is treated by inclusion of stochastic terms.}. 

Reduction of noise does not necessary solve the problem of diverging stochastic trajectories, which can be avoided by using discrete distributions as follows. Whenever the phase-space variable becomes too large we expand the corresponding kernel in terms of an equivalent discrete distribution. Then we randomly choose (according to the obtained discrete distribution) one of the discrete phase-point operators (kernels) and continue the simulation with the chosen kernel (for details see \aref{App2} and \aref{App3}). For a large number of simulated trajectories this procedure converges to the exact result. 

\section{Example: dissipative long-range transverse Ising chain}
\begin{widetext}
\begin{figure*}[h!!]
\vskip 0.5truecm
\includegraphics[width=0.24\textwidth]{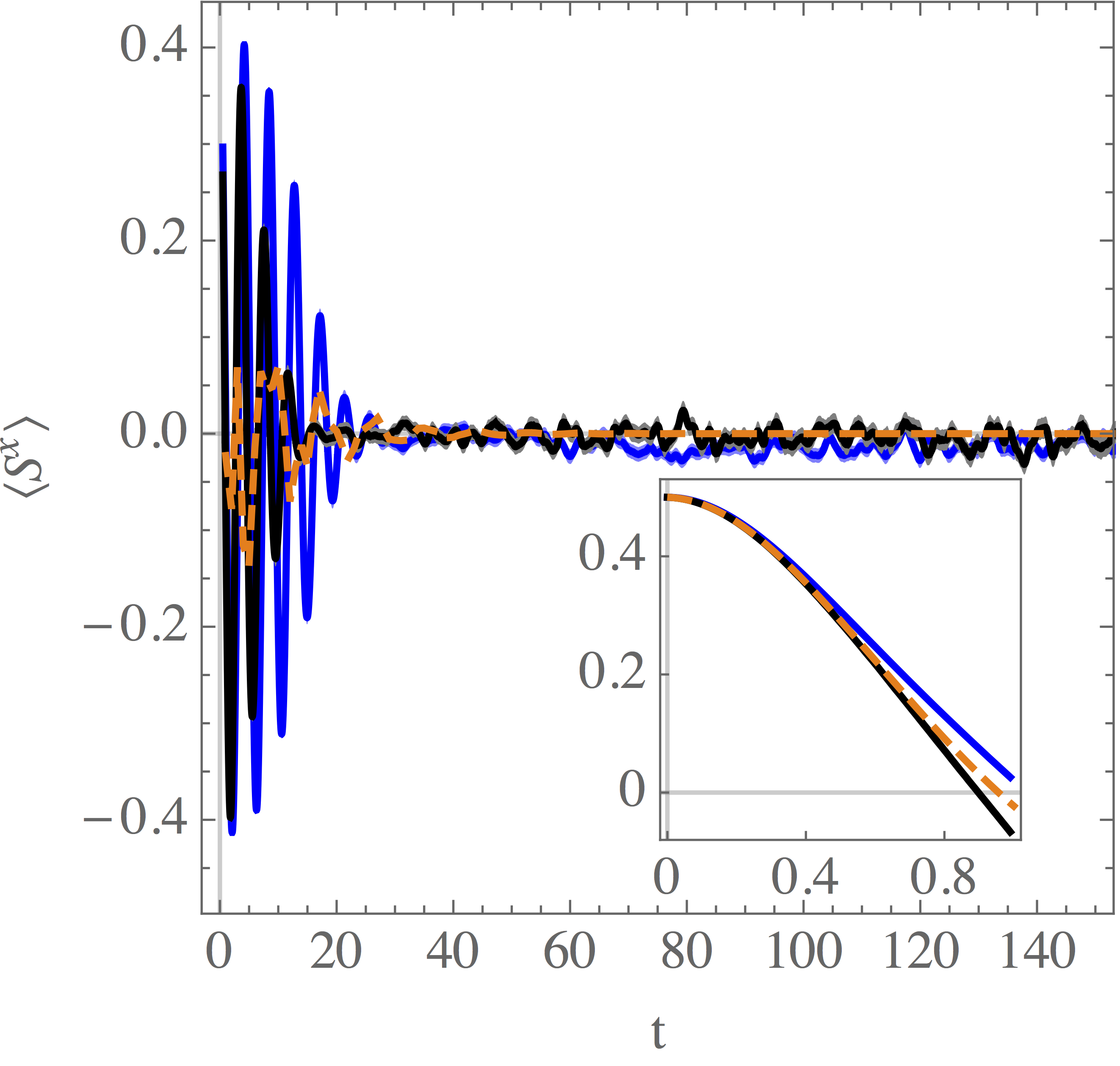}~~\includegraphics[width=0.24\textwidth]{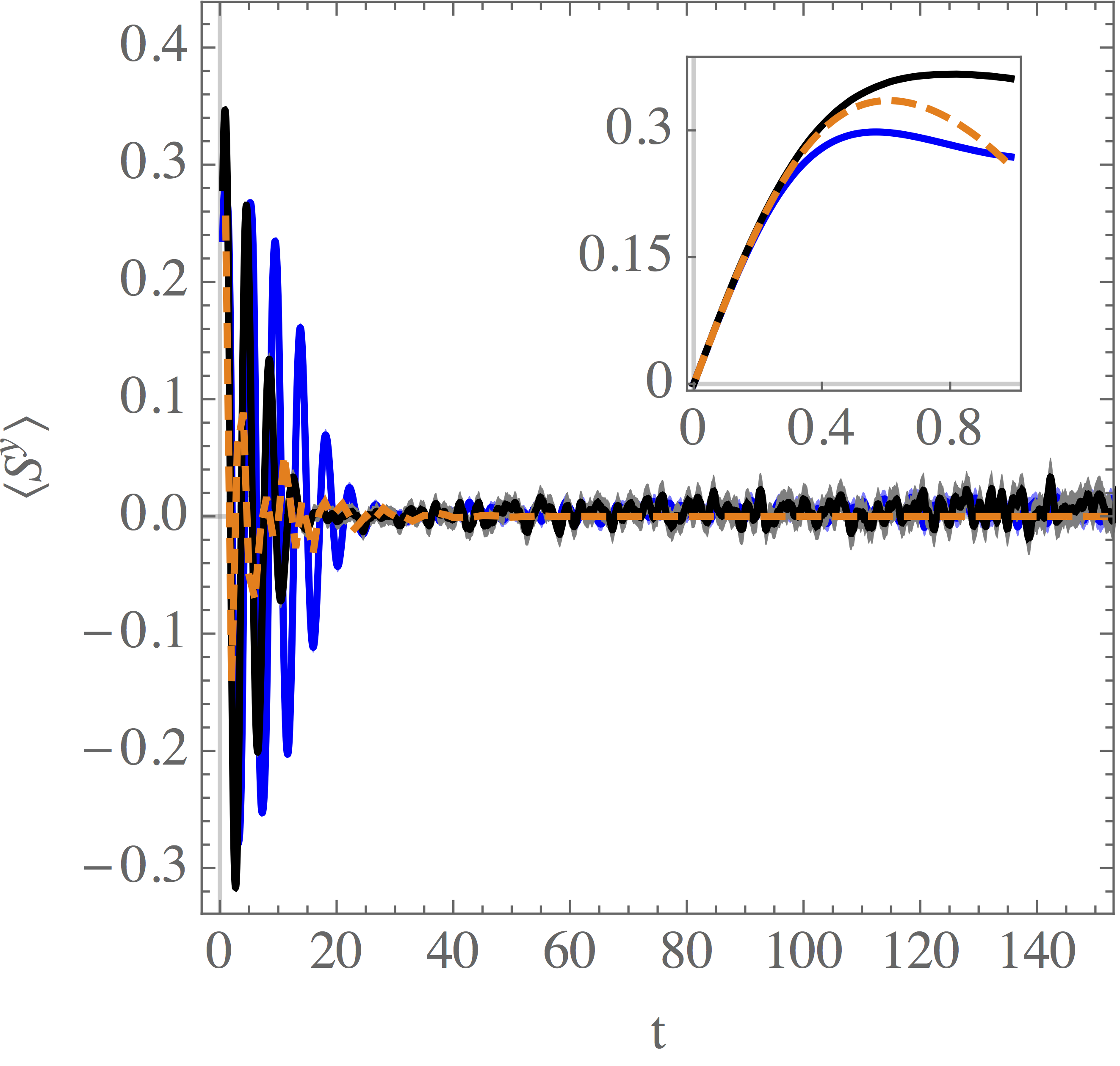}~~\includegraphics[width=0.24\textwidth]{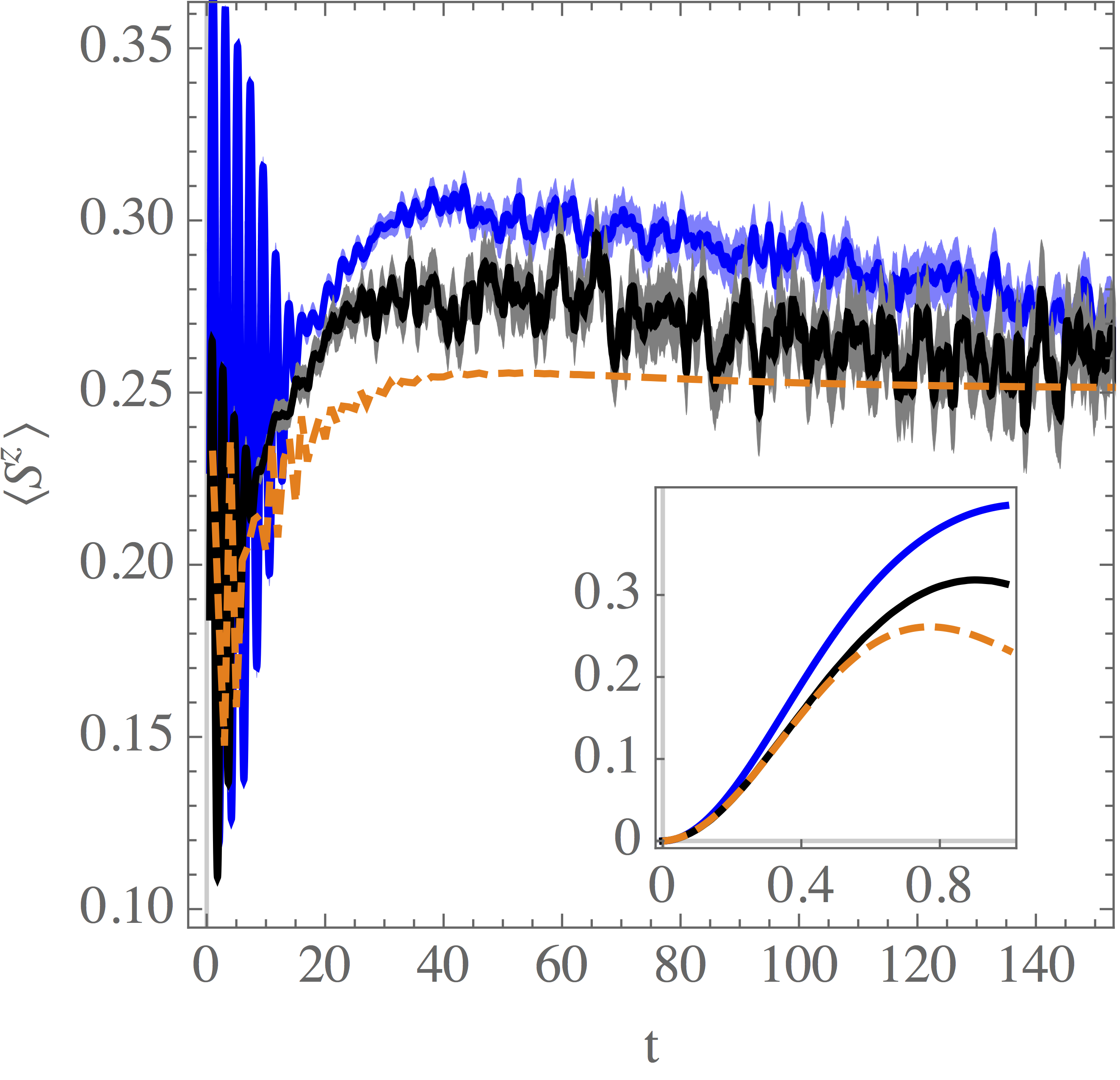}\\
\includegraphics[width=0.24\textwidth]{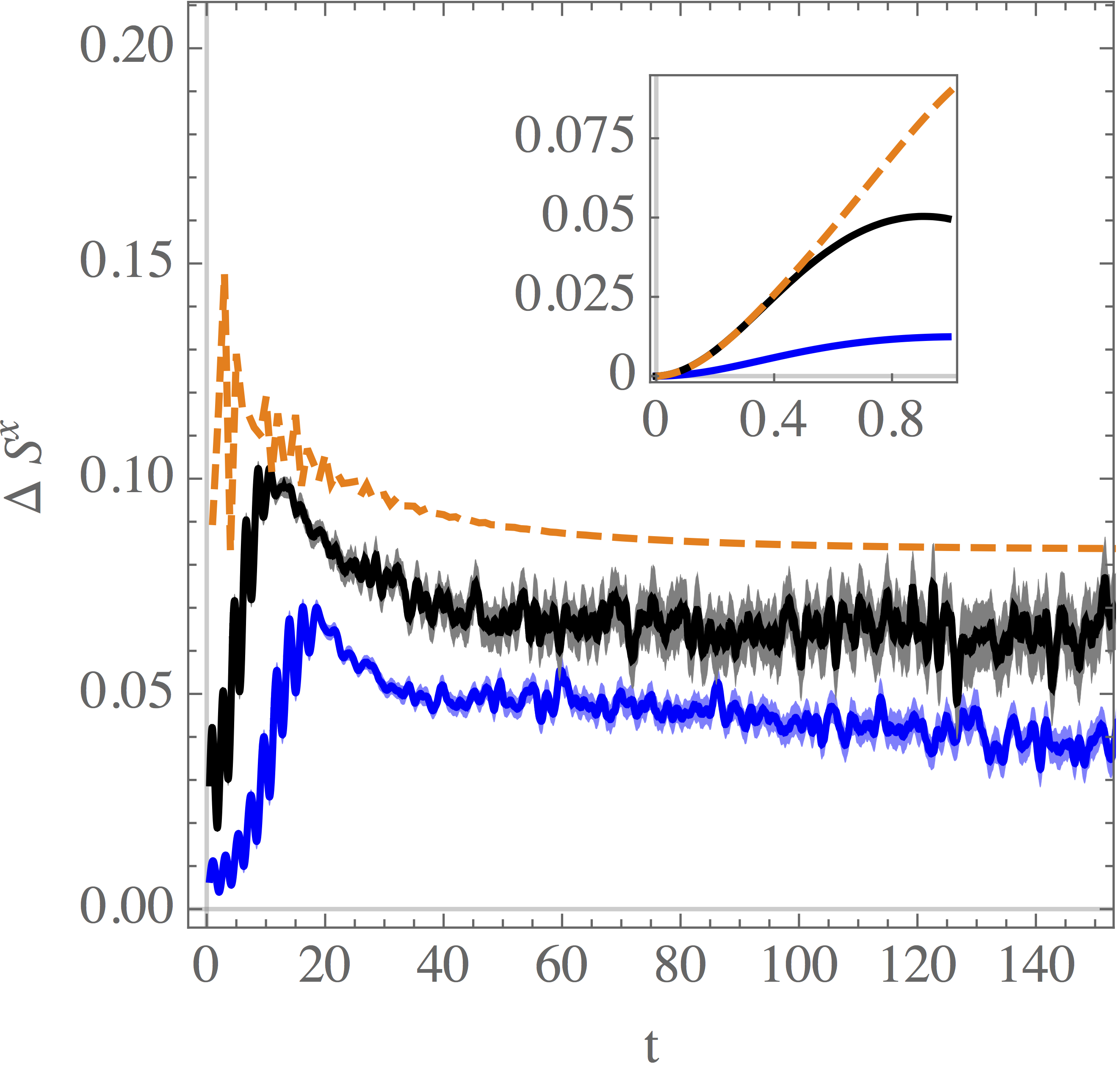}~~\includegraphics[width=0.24\textwidth]{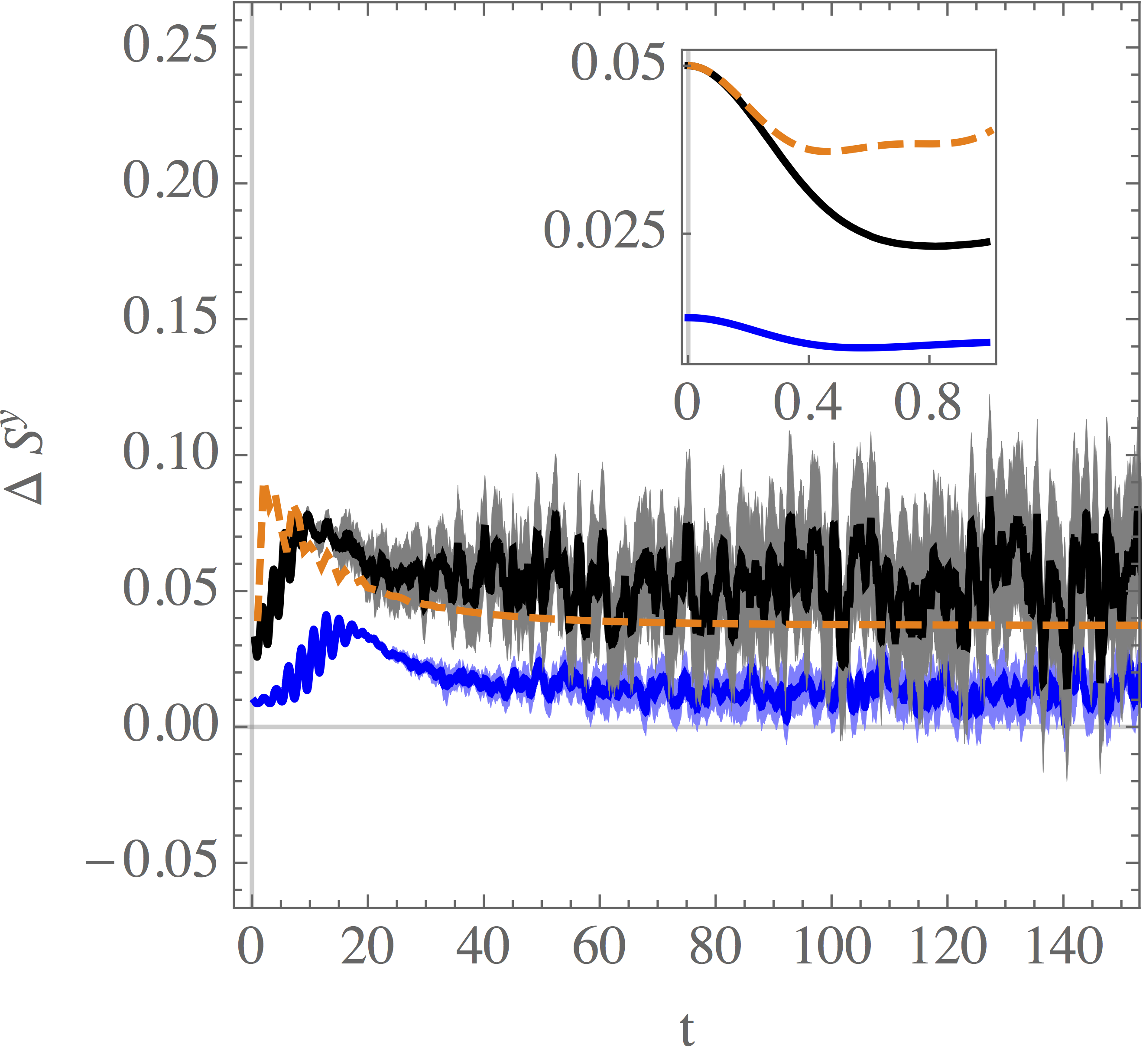}~~\includegraphics[width=0.24\textwidth]{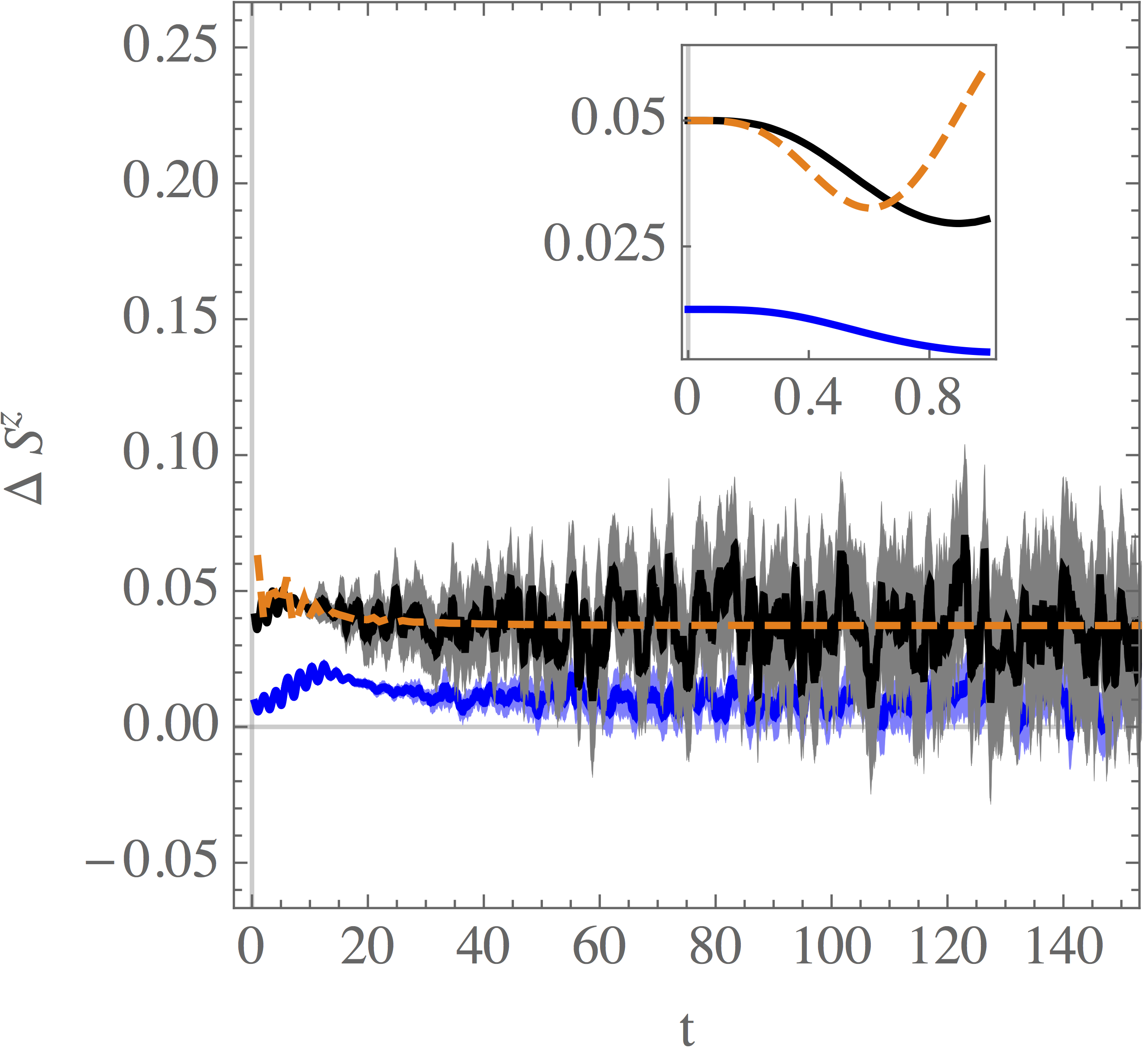}
\caption{Top figures show the average spin per particle in $x,y,z$ direction, $S^{\rm \alpha}=\frac{1}{2n}\sum_{k=1}^{n}\sigma^{\alpha}_j$, $\alpha={\rm x,y,z}$. Bottom figures show the two point correlations $\Delta S^{\rm \alpha}=\ave{S^{\rm \alpha}S^{\rm \alpha}}-\ave{S^{\rm \alpha}}^2$. The orange dashed line represents the exact result for $n=5$, the black and blue lines show the stochastic simulation result for $n=5$ and $n=20$, respectively. The grey/light blue regions denote the statistical error of the stochastic simulation with  $10^3$ trajectories.  The initial state is a product state with all spins pointing in the $x$ direction. Insets show the short time dynamics.  Parameters used in the simulation are: $\alpha=1.5,~h=1,~\gamma_1=0.2,~ \gamma_2=0.02,~ \gamma_3=0.1,~ \gamma_4=0.05,~ \gamma_{\rm D}=0.001$.} 
\label{fig1}
\end{figure*}
\end{widetext}

As an example of the discrete phase space projection method described in the previous Section we consider a long-range dissipative Ising chain in a transverse field with the Hamiltonian
\begin{align}
H=\sum_{j,k=1}^n\frac{J(\alpha,n)}{2|j-k|^\alpha}\sx_j\sx_k +h\sum_j\sz_j,
\end{align}
boundary Lindblad operators $L_1=\sqrt{\gamma_1}\sigma^+_1,~L_2=\sqrt{\gamma_2}\sigma^-_{1},~L_3=\sqrt{\gamma_3}\sigma^+_{n},~L_4=\sqrt{\gamma_4}\sigma^+_{n}$, and bulk dephasing $L_{4+j}=\sqrt{\gamma_{\rm D}}\sz_j$, $j=1,2,\ldots n$. We are using the Kac normalization $J(\alpha,n)=(\sum_{j=1}^n j^{-\alpha})^{-1}$. The time evolution of the density matrix is given by the Lindblad equation
\begin{align}
\ddt \rho=-\ii [H,\rho]+\sum_{\mu=1}^{n+4}L_\mu\rho L_\mu^\dag-\frac{1}{2}\{L_\mu^\dag L_\mu,\rho\}
\end{align}
and can be expressed as a partial differential equation on the extended phase space \cite{GZ04} (see \aref{App1}).  This problem can not be solved exactly by any known method to treat open quantum systems \cite{Pro08,ZP10,Zni10,Zun14,IZ14} and it is hard to simulate with existing numerical methods \cite{ZP09}. Usually we are interested in the steady state of open quantum systems. In order to estimate the long-time behaviour we consider the interaction semiclassically, whereas the dissipation is treated fully quantum mechanically. For small system sizes ($n=5$) we compare the results obtained by the proposed method with exact results and observe good agreement for short and long times. It is surprising that semiclassical approximation of the interaction provides an accurate estimation of the long-time behaviour even in the strongly interacting regime (see \fref{fig1}). We also compute the long-time averages for a larger system with $n=20$.  Parameters used in presented simulations are: $\alpha=1.5,~h=1,~\gamma_1=0.2,~ \gamma_2=0.02,~ \gamma_3=0.1,~ \gamma_4=0.05,~ \gamma_{\rm D}=0.001$. 

\section{Conclusions and discussion}We have shown that the phase-point operators defining discrete quasiprobability distributions are equivalent to the continuous quantisation kernel evaluated at spacial points of the phase space. This identification justifies use of continuous phase-space methods on discrete phase spaces. Further, we demonstrated that discrete phase-space sampling can be used to treat the problem of diverging trajectories. We applied the procedure on a dissipative transverse Ising chain and calculated the long-time expectation values of local observables up to system size $n=20$. We found that long-time behaviour of local observables is well described by a semiclasical  approximation of the interaction even in the strongly interacting regime. It is an interesting question if this generalises to other models. For example to the integrable XXZ chain, where the steady state is described by a quasilocal operator \cite{Pro11}. The discrete phase space can be regarded as a special basis in the space of operators. Thus, the established connection between discrete and continuous phase space may enable a construction of mixed Hilbert-space--phase-space methods for simulation of many-body quantum systems. From a more theoretical point of view it would be interesting to explore its implications on the problem of mutually unbiased bases in non-prime number dimensions.

\section*{Acknowledgments}
The author thanks Felipe Barra for stimulating discussions. This work was supported by Chilean FONDECYT project 3130495.

\bibliography{Bibliography}
\appendix
\widetext

\section{Derivation of the Fokker-Planck equation for the positive-P function}
\label{App1}
In this Appendix we derive a partial-differential equation of the problem considered in paper. The Hamiltonian is given by
\begin{align}
H=\sum_{j,k=1}^n\frac{J}{2|j-k|^\alpha}\sx_j\sx_k +h\sum_j\sz_j,
\end{align}
the boundary Lindblad operators are $L_1=\sqrt{\gamma_1}\sigma^+_1,~L_2=\sqrt{\gamma_2}\sigma^-_{1},~L_3=\sqrt{\gamma_3}\sigma^+_{n},~L_4=\sqrt{\gamma_4}\sigma^+_{n}$ and the bulk dephasing Lindblad operators are $L_{j+4}=\sqrt{\gamma_D}\sz_j,~j=1,\ldots n$. We are using the Kac normalization $J(\alpha,n)=(\sum_{j=1}^n j^{-\alpha})^{-1}$. Time evolution of the density matrix is determined by the Lindblad equation 
\begin{align}
\ddt \rho=-\ii [H,\rho]+\sum_{\mu=1}^{n+4}L_\mu\rho L_\mu^\dag-\frac{1}{2}\{L_\mu^\dag L_\mu,\rho\}.
\label{eqa:lind}
\end{align}
In order to arrive at the phase space formulation of the equation \eref{eqa:lind} we follow \cite{BD08}. First we define unnormalized $SU(2)$ coherent states for $j-th$ spin as
\begin{align}
\kket{\psi_j}=\ket{0}+\psi_j\ket{1}.
\end{align}
A coherent state for the complete system is then 
\begin{align}
\kket{\ul{\psi}}=\kket{\psi_1}\otimes\kket{\psi_2}\otimes\ldots\kket{\psi_n}.
\end{align}
The action of pauli operators $\sigma_j^{\alpha}$, $\alpha=\mathrm{x,y,z}$ on a coherent state can be expressed in terms of partial derivatives as 
\begin{align}
\sx_j\kket{\ul{\psi}}=&\left(\psi_j+(1-\psi_j^2)\partial_{\psi_j}\right)    \kket{\ul{\psi}},\\ \nonumber
\sy_j\kket{\ul{\psi}}=&\ii \left(-\psi_j+(1+\psi_j^2)\partial_{\psi_j} \right)    \kket{\ul{\psi}},\\ \nonumber
\sz_j\kket{\ul{\psi}}=&\left( 1-2\psi_j\partial_{\psi_j} \right)    \kket{\ul{\psi}}.
\end{align}
We can expand the density matrix in terms of many-body coherent states as
\begin{align}
\rho=&\int_{\mc{M}}\dd\mu(\ul{\psi})\dd\mu(\ul{\phi}) P(\ul{\psi},\ul{\phi})\Lambda(\ul{\psi},\ul{\phi}),\\ \nonumber
\Lambda(\ul{\psi},\ul{\phi})&=\frac{\kket{\ul{\psi}}\bbra{\ul{\bar{\phi}}}}{\brakket{\ul{\bar{\phi}}}{\ul{\psi}}},
\end{align}
where the bar $\bar{\bullet}$ denotes complex conjugation. In order to derive the partial differential equation for the density matrix we have to specify the action of the Pauli operators on the qunatization kernel $\Lambda(\ul{\psi},\ul{\phi})$
\begin{align}
\label{eqa:pauliL}
\sx_j\Lambda(\ul{\psi},\ul{\phi})&=\left(  \frac{\psi_j+\phi_j}{1+\psi_j\phi_j}+(1-\psi_j^2)\partial_{\psi_j}  \right)\Lambda(\ul{\psi},\ul{\phi}),\\\nonumber
\Lambda(\ul{\psi},\ul{\phi})\sx_j&=\left(  \frac{\psi_j+\phi_j}{1+\psi_j\phi_j}+(1-\phi_j^2)\partial_{\phi_j}  \right)\Lambda(\ul{\psi},\ul{\phi}),\\\nonumber
\sy_j\Lambda(\ul{\psi},\ul{\phi})&=\ii \left( - \frac{\psi_j+\phi_j}{1+\psi_j\phi_j}+(1+\psi_j^2)\partial_{\psi_j} \right)\Lambda(\ul{\psi},\ul{\phi}),\\\nonumber
\Lambda(\ul{\psi},\ul{\phi})\sy_j&=\ii \left( - \frac{\psi_j+\phi_j}{1+\psi_j\phi_j}+(1+\phi_j^2)\partial_{\phi_j} \right)\Lambda(\ul{\psi},\ul{\phi}),\\\nonumber
\sz_j\Lambda(\ul{\psi},\ul{\phi})&=-\left( \frac{1-\psi_j\phi_j}{1+\psi_j\phi_j}-2\psi_j\partial_{\psi_j} \right)\Lambda(\ul{\psi},\ul{\phi}),\\\nonumber
\Lambda(\ul{\psi},\ul{\phi})\sz_j&=-\left( \frac{1-\psi_j\phi_j}{1+\psi_j\phi_j}-2\phi_j\partial_{\phi_j}  \right)\Lambda(\ul{\psi},\ul{\phi}).
\end{align}
Applying above equations \eref{eqa:pauliL} to the commutator of the Hamiltonian with the quantisation kernel $\Lambda$ (in the following we use for brevity $\Lambda$ instead of $\Lambda(\ul{\Psi},\ul{\Phi})$ and $J$ instead of $J(\alpha,n)$) we find 
%\begin{widetext}
\begin{align}
\label{eqa:hamL}
-\ii[H,\Lambda]=&-\ii\sum_{j,k=1}^n\frac{J}{2|j-k|^\alpha}\left(\frac{\psi_j+\phi_j}{1+\psi_j\phi_j} \left( (1-\psi_{k}^2)\partial_{\psi_{k}}-(1-\phi_{k}^2)\partial_{\phi_{k}} \right)+\frac{\psi_{k}+\phi_{k}}{1+\psi_{k}\phi_{k}} \left( (1-\psi_{j}^2)\partial_{\psi_{j}}-(1-\phi_{j}^2)\partial_{\phi_{j}} \right) \right)\Lambda \\ \nonumber
&-\ii\sum_{j,k=1}^n\frac{J}{2|j-k|^\alpha}\left((1-\psi_j^2)(1-\psi_{k}^2)\partial_{\psi_j}\partial_{\psi_{k}}-(1-\phi_j^2)(1-\phi_{k}^2)\partial_{\phi_j}\partial_{\phi_{k}}  \right)\Lambda\\ \nonumber
&-2\ii h\sum_{j-1}^n\left( \phi_j\partial_{\phi_j}-\psi_j\partial_{\psi_j} \right)\Lambda.
\end{align}
Similarly we compute the action of the dissipators on the quantization kernel
\begin{align}
\label{eqa:dissL}
\sigma^+_j\Lambda\sigma^-_j-\frac{1}{2}\{\sigma^-_j\sigma^+_j,\Lambda\}&=\left(\frac{3+\psi_j\phi_j}{2+2\psi_j\phi_j}(\psi_j\partial_{\psi_j}+\phi_j\partial_{\phi_j})+\partial_{\psi_j}\partial_{\phi_j}\right)\Lambda,\\ \nonumber
\sigma^-_j\Lambda\sigma^+_j-\frac{1}{2}\{\sigma^+_j\sigma^-_j,\Lambda\}&=\left(\frac{1+3\psi_j\phi_j}{2+2\psi_j\phi_j}(\psi_j\partial_{\psi_j}+\phi_j\partial_{\phi_j})+\psi_j^2\phi_j^2\partial_{\psi_j}\partial_{\phi_j}\right)\Lambda,\\ \nonumber
\sigma^z_j\Lambda\sigma^z_j-\Lambda&=\left(\frac{-2+2\psi_j\phi_j}{1+1\psi_j\phi_j}(\psi_j\partial_{\psi_j}+\phi_j\partial_{\phi_j})+4\psi_j\phi_j\partial_{\psi_j}\partial_{\phi_j}\right)\Lambda.
\end{align}
%\end{widetext}
In order to further simplify the notation we define a vector of phase space variables $\ul{z}=(\psi_1,\ldots\psi_n,\phi_1,\ldots\phi_n)$ and the partial derivatives $\partial_j\equiv\partial_{\psi_j}$, $\partial_{n+j}\equiv\partial_{\phi_{j}}$. Using the equations \eref{eqa:hamL} and \eref{eqa:dissL} we obtain the following equation for the density matrix
\begin{align}
\label{eqa:partL}
\ddt \rho&=\int_{\mc{M}}\dd\mu(\ul{\psi})\dd\mu(\ul{\phi})\mc{L}_\Lambda \Lambda,\\ \nonumber
\mc{L}_{\Lambda}&=\sum_{j=1}^{2n}A_j\partial_j+\frac{1}{2}\sum_{j,k=1}^{2n}D_{j,k}\partial_j\partial_k,
\end{align}
where the non-zero elements of the drift vector $A$ and the diffusion matrix $D$ are given by
%\begin{widetext}
\begin{align}
A_j=&\delta_{j,1}\left(\gamma_1 \frac{3+\psi_j\phi_j}{2+2\psi_j\phi_j}\psi_j+\gamma_2\frac{1+3\psi_j\phi_j}{2+2\psi_j\phi_j}\psi_j \right)+\delta_{j,n}\left(\gamma_3 \frac{3+\psi_j\phi_j}{2+2\psi_j\phi_j}\psi_j+\gamma_4\frac{1+3\psi_j\phi_j}{2+2\psi_j\phi_j}\psi_j \right)+\gamma_D\frac{-2+2\psi_j\phi_j}{1+1\psi_j\phi_j}\psi_j,\\ \nonumber
&-2\ii h\psi_j-\ii\sum_{k=1;~k\neq j}^n\frac{J}{|j-k|^\alpha}\left(\frac{\psi_{k}+\phi_{j+1}}{1+\psi_{k}\phi_{k}}(1+\psi_{k}^2)\right),\\ \nonumber
A_{n+j}=&\delta_{j,1}\left(\gamma_1 \frac{3+\psi_j\phi_j}{2+2\psi_j\phi_j}\phi_j+\gamma_2\frac{1+3\psi_j\phi_j}{2+2\psi_j\phi_j}\phi_j \right)+\delta_{j,n}\left(\gamma_3 \frac{3+\psi_j\phi_j}{2+2\psi_j\phi_j}\phi_j+\gamma_4\frac{1+3\psi_j\phi_j}{2+2\psi_j\phi_j}\phi_j \right)+\gamma_D\frac{-2+2\psi_j\phi_j}{1+1\psi_j\phi_j}\phi_j,\\ \nonumber
&+\ii2h\phi_j+\ii\sum_{k=1;~k\neq j}^n\frac{J}{2|j-k|^\alpha} \left(\frac{\psi_{k}+\phi_{k}}{1+\psi_{k}\phi_{k}}(1+\phi_{k}^2)\right),\\ \nonumber
D_{j+n,j}=&D_{j,j+n}=\delta_{j,1}(\gamma_1+\gamma_2\psi_j^2\phi_j^2) \delta_{j,n}(\gamma_3+\gamma_4\psi_j^2\phi_j^2),+4\gamma_D\psi_j\phi_j,\\ \nonumber
D_{j,j+1}=&D_{j+1,j}=-\frac{J\ii}{2|j-k|^\alpha} (1-\psi_j^2)(1-\psi_{j+1}^2),\\ \nonumber
D_{n+j,n+j+1}=&D_{n+j+1,n+j}=-\frac{J\ii}{2|j-k|^\alpha} (1-\phi_j^2)(1-\phi_{j+1}^2).
\end{align}
%\end{widetext}
By partial integration of the equation \eref{eqa:partL} and assuming that boundary terms vanish we obtain the following partial differential equation for the positive-P function
\begin{align}
\label{eqa:partP}
\ddt P(\ul{z})&=\left(-\sum_{j=1}^{2n}\partial_jA_j+\frac{1}{2}\sum_{j,k=1}^{2n}\partial_j\partial_kD_{j,k}\right)P(\ul{z}).
\end{align}
This equation is a Fokker-Planck equation which can be simulated by rewriting it to a stochastic differential equation  (see e.g. \cite{BD08}).
\section{Discrete phase-space solution for divergent trajectories}
\label{App2}
In this section a simple example of discrete phase-space projection is discussed.  We will demonstrate that the diverging trajectories are a consequence of a particular choice of the phase space and could in principle be avoided by using a redundant representation of the phase space. We consider the SU(2) case (i.e. a spin 1/2). In this case the phase space pertaining to the positive-P function is $\mathds{C}^2$, which can be conveniently represented by using stereographic projection to two spheres. To each point in the phase space corresponds a kernel $\Lambda(z,v)=\kket{z}\bbra{\bar{v}}/\brakket{\bar{v}}{z}$. Since this is an operator in the Hilbert space it can be expanded using the discrete phase space points as $\Lambda(z,v)=\sum_{\alpha,\beta,\gamma,\delta}p_{\alpha,\beta,\gamma,\delta}\Lambda(z_{\alpha,\beta},z_{\gamma,\delta})$, with $z_{\alpha,\beta}$ defined in Section \ref{sec:geom} and $p_{\alpha,\beta,\gamma,\delta}$  positive real numbers summing to one. If at some point in the stochastic evolution $|z_j(t)|$ or $|v_j(t)|$ crosses a threshold $z_{max}$ we calculate the discrete probability distribution $p^{(j)}_{\alpha,\beta,\gamma,\delta}$ corresponding to the phase-space variables $z_j(t)$, $v_j(t)$ at time $t$ and randomly choose one discrete phase space point $(z_{\alpha,\beta},v_{\gamma,\delta})$ according to the distribution $p_{\alpha,\beta,\gamma,\delta}$. We interpret the chosen kernel (phase space point $(z_j(t), v_j(t))\rightarrow(z_{\alpha,\beta},v_{\gamma,\delta})$) as one corresponding to a continuous-phase space distribution and continue the evolution from this point. Phase-space variables corresponding to particles at sites $k\neq j$ do not change.

\begin{figure}[h!!]
\vskip 0.5truecm
\includegraphics[width=0.25\textwidth]{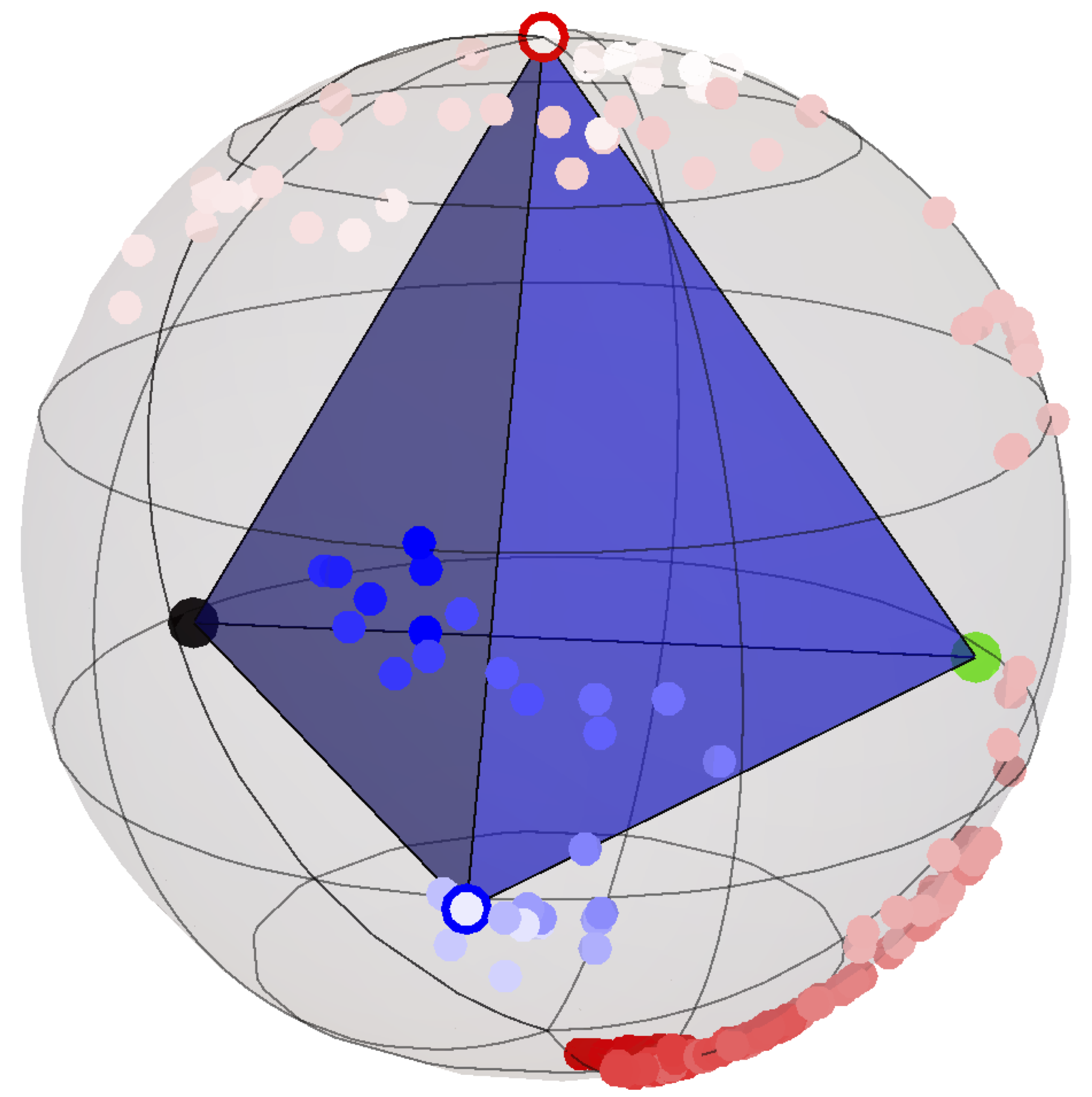}
\caption{Demonstration of the algorithm on a particular trajectory. The red and blue points denote the stochastic trajectory before and after the jump, respectively. The brightness of the color denotes how far in the past is the trajectory (the initial point and the point after the jump are white). The dark red points are close to the $z_{max}=10\sqrt{2}$, when the trajectory crosses this value it is projected to the blue discrete phase-space point. After that the stochastic evolution continues.}
\label{fig3}
\end{figure}

We found numerically that in all cases $p_{\alpha,\beta,\gamma,\delta}$ are positive and sum to one.  In \fref{fig3} we show an example trajectory corresponding to the variable $z_1(t)$ of the first spin. Since we fixed the highest weight representation the condition $|z(t)|<z_{max}$ means that the spin in the $z$ direction cannot be smaller than $\cos(\theta_{max}/2)$. If at some time the trajectory of the phase-space variable crosses the critical circle on the sphere we project the phase-space point at the crossing to one point on the sphere corresponding to its discrete quasi-probability distribution. From this picture it is clear that the divergence of the trajectory is related to the inability of the chosen kernel to represent all states in the Hilbert space by coherent states corresponding to the phase-space region with $|z|<z_{max}$.

\begin{figure}[h!!]
\vskip 0.5truecm
\includegraphics[width=0.4\textwidth]{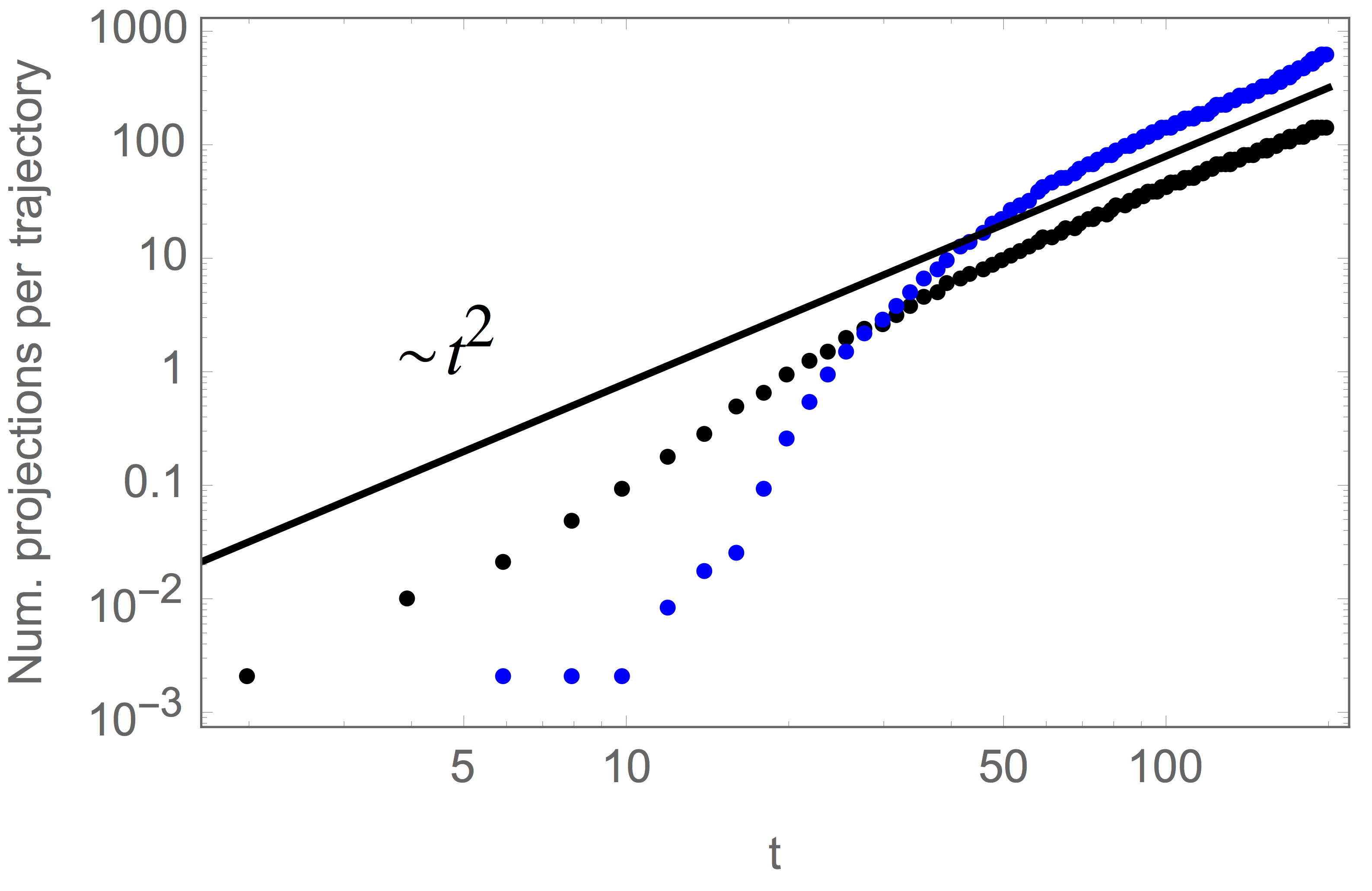} 
\caption{Average number of discrete projections per trajectory until time $t$. The black, blue dots correspond to $n=5,20$. The full line serves as a guide to the eye and is proportional to $t^{2}$.}
\label{fig4}
\end{figure}

In \fref{fig4} we show the average number of projections (jumps to the discrete distribution) per trajectory for system sizes $n=5,~20$.

\section{Calculation of local expectation values}
\label{App3}
Once we have an ensemble of stochastic trajectories we calculate the expectation values of local observables (e.g. on site $j$) by stochastic averages 
\begin{align}
\ave{O_j}=\trb{\rho O_j}=\ave{O(\psi_j,\phi_j)}_{\rm stoch},
\end{align}
where $\ave{\bullet}_{\rm stoch}$ denotes an average over stochastic trajectories and
\begin{align}
O(\psi_j,\phi_j)=\trb{O\Lambda(\psi_j,\phi_j)}.
\end{align}
In the spin 1/2 case the local observables are given by 
\begin{align}
\sx(\psi,\phi)=\frac{\psi+\phi}{1+\phi\psi},\quad
\sy(\psi,\phi)=\ii\frac{\psi-\phi}{1+\phi\psi},\quad 
\sz(\psi,\phi)=\frac{1-\psi\phi}{1+\phi\psi}.
\end{align}
Since these functions have poles at $\phi\psi=-1$ we perform discrete projections also when $|\phi_j\psi_j-1|<\epsilon$  (in the simulations we take $\epsilon=0.1$). 

\end{document}